\documentclass[letterpaper]{article} 

\usepackage{aaai2027}

\usepackage[hyphens]{url}  
\usepackage{graphicx} 
\usepackage{natbib}  
\usepackage{caption} 
\usepackage{algorithm}
\usepackage{algorithmic}
\usepackage{booktabs}
\usepackage{multirow}
\usepackage{amsmath}
\usepackage{makecell}
\usepackage{amssymb}

\usepackage[table]{xcolor} 

\title{Hear, Invoke, and Understand: A Skill-Calling Multimodal Agent for Large Audio Language Models}

\author{
    Yuwen Wang\textsuperscript{\rm 1,\rm 2}\equalcontrib,
    Tian-Hao Zhang\textsuperscript{\rm 2}\equalcontrib,
    Minghao Cai\textsuperscript{\rm 1},
    Yilin Ren\textsuperscript{\rm 1},
    Ziyang Jiang\textsuperscript{\rm 1},\\
    Xin Wang\textsuperscript{\rm 2},
    Zhichao Wang\textsuperscript{\rm 2},
    Zhou Pan\textsuperscript{\rm 2},
    Kun Zhan\textsuperscript{\rm 2},
    Xinyuan Qian\textsuperscript{\rm 1}
}

\affiliations{
    \textsuperscript{\rm 1}
    University of Science and Technology Beijing, China\\
    \textsuperscript{\rm 2}
    Li Auto Inc., Beijing, China\\
}

\begin{document}

\nocopyright

\maketitle

\begin{abstract}
Complex acoustic problems may require models to perform acoustic operations, interact with external tools and reason over the resulting textual or processed-audio observations rather than answer directly from a fixed audio input.
We study such problems as tool-interactive audio reasoning and develop \textit{SpeechAgent-R}, an audio agent that coordinates its intrinsic multimodal understanding with external skills and tools.
To support this capability, we construct \textit{HIU-Corpus}, comprising 65,492 interaction trajectories and 507.6 hours of audio across 24 tasks, 8 skills and 9 tools.
\textit{SpeechAgent-R} first learns structured interaction behaviors through trajectory-based supervised fine-tuning and then improves its decisions through multi-turn reinforcement learning. 
We further introduce \textit{HIU-Bench} to jointly evaluate task performance, interaction quality and generalization to diverse task settings. 
It contains 1,395 samples across 56 tasks, including in-distribution (ID) and out-of-distribution (OOD) splits with substantial shifts in tool usage and workflow composition.
\textit{SpeechAgent-R} achieves 84.17 on ID tasks and 70.94 on OOD tasks, improving over the base model under the same agent harness by 15.40 and 14.23 points.
These results demonstrate that learning skill and tool coordination improves audio agents' ability to handle diverse task settings and adaptive tool interactions.
\end{abstract}

\section{Introduction}

Recent advances in large audio language models (LALMs) have improved their ability to understand diverse audio content \cite{qwen2-audio,step-audio-r1}. 
Meanwhile, benchmarks such as MMAU and MMAR \cite{mmau,mmar} have introduced more challenging problems to evaluate these capabilities. 
However, two important limitations remain. 
First, existing LALMs mainly rely on internal inference and parametric knowledge acquired during training. 
When a task requires acoustic operations or reasoning procedures beyond their learned capabilities, they lack mechanisms to acquire such capabilities at inference time.
Second, unlike text-based agents that can learn new skills from task instructions and tool descriptions, current multimodal models generally lack mechanisms to extend their capabilities through such instructions at inference time.
This ability to acquire and apply new skills remains underexplored in audio understanding.

\begin{figure}[t]
\centering
\includegraphics[width=1.0\columnwidth]{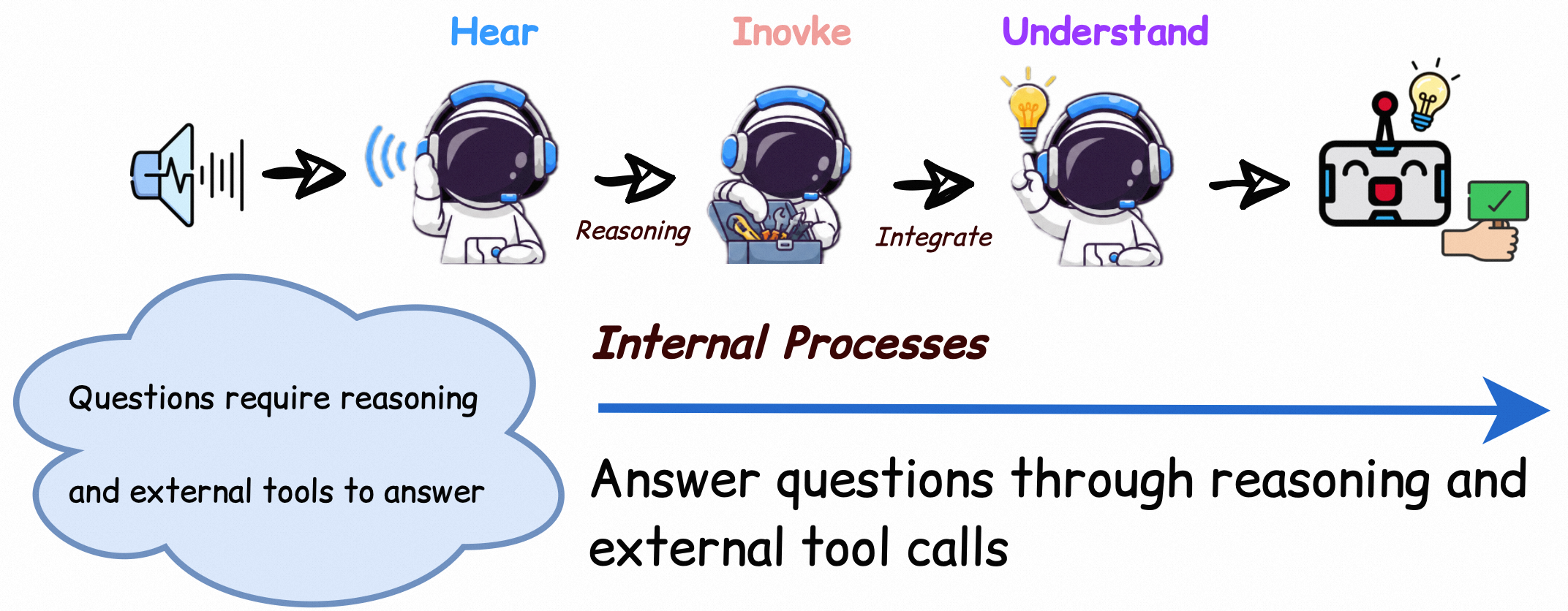}
\caption{Overview of the \textit{Hear-Invoke-Understand} framework. SpeechAgent-R hears the acoustic scene, invokes required skills through task reasoning, and understands the problem by integrating the returned evidence.}
\label{figure1}
\end{figure}

These limitations are particularly consequential in complex acoustic scenes, where task-relevant evidence must be actively acquired and integrated \cite{sussman}. 
For example, recovering the speech of a target speaker from a spatial recording may require localization, speaker identification, speech extraction and semantic interpretation, with each operation depending on preceding results. 
Recent audio agents have explored external tools, but do not yet support this adaptive process: 
AudioToolAgent \cite{audiotoolagent} uses a text-based LLM to coordinate audio models but lacks native audio reasoning; 
AuTAgent \cite{autagent} learns tool selection through reinforcement learning but does not compose tools based on intermediate outcomes; 
Echo \cite{echo} enables audio re-listening but remains restricted to the original input; 
and EChO-Agent \cite{echoagent}improves evidence organization within predefined interaction workflows. 
Consequently, existing approaches remain limited in their ability to adapt tool usage and compose workflows for previously unseen tasks and newly introduced skills.
The central challenge is therefore to enable audio models to reason over multi-step interactions, coordinate diverse tools based on skill descriptions, and integrate tool-generated audio and textual observations into coherent decision making.

To address this challenge, we formulate complex audio problem solving as an iterative \textit{Hear–Invoke–Understand (HIU)} process, as illustrated in Figure~\ref{figure1}. 
Given an audio query, the model first \textit{hears} the input audio with its native perception, \textit{invokes} appropriate tools when needed and \textit{understands} the resulting textual or audio observations before deciding the next action.
Learning this process requires complete execution trajectories, which existing audio datasets do not provide. 
We therefore construct \textit{HIU-Corpus}, comprising 65,492 samples and 507.6 hours of audio across 24 tasks, 8 skills and 9 tools. 
Using this corpus, we develop \textit{SpeechAgent-R}, a skill-calling multimodal agent trained with trajectory-based supervised fine-tuning (SFT) and multi-turn reinforcement learning (RL) to learn skill selection and tool coordination. 
We further introduce \textit{HIU-Bench} to evaluate task performance, interaction quality and generalization across diverse task and skill settings.
It contains 1,395 samples spanning 56 tasks and 26 tools, with ID and OOD subsets sharing only two tools and no annotated workflows. 
SpeechAgent-R achieves an overall score of 80.05, outperforming its base model under the same agent harness by 15.04 points and demonstrating the potential of LALMs to move beyond direct audio understanding toward interactive problem solving with external skills.
Our contributions are:
\begin{itemize}
    \item We formulate complex audio problem solving as an adaptive multi-step interaction in which audio agents combine external tools with their multimodal understanding.
    \item We construct \textit{HIU-Corpus} and develop \textit{SpeechAgent-R} with trajectory-based SFT and multi-turn RL, enabling the model to learn skill-based tool coordination.
    \item We introduce \textit{HIU-Bench} to evaluate audio agents from task performance, interaction quality and generalization to unseen configurations.
\end{itemize}
\section{Related Work}

\subsection{LALMs and Audio Reasoning Benchmarks}

LALMs have evolved toward unified understanding of speech, environmental sounds and music. 
Models such as Qwen2-Audio, Kimi-Audio and Qwen3-Omni support general audio understanding and instruction following \cite{qwen2-audio,kimi-audio,qwen3-omni}, while Audio-Reasoner, Audio-Cogito and Step-Audio-R1 further improve complex reasoning through audio-grounded reasoning data and post-training \cite{audioreasoner,audiocogito,step-audio-r1}. 
These advances improve reasoning over fixed audio inputs, yet the models still operate on the original input alone, without acquiring additional evidence through acoustic operations or incorporating tool-generated observations into subsequent reasoning.

Audio benchmarks have accordingly expanded from general understanding to complex reasoning. 
AIR-Bench and AudioBench evaluate the understanding of speech, sounds and music \cite{airbench,audiobench}, while MMAU, MMAR and MMAU-Pro introduce professional knowledge, multi-step reasoning, multiple audio inputs and spatial audio problems \cite{mmau,mmar,mmaupro}. 
MSU-Bench further focuses on speaker-centric understanding in multi-speaker conversations \cite{msubench}. 
Despite increasingly challenging tasks and more detailed reasoning evaluation, these benchmarks primarily evaluate models’ ability to reason over provided audio inputs.
They do not evaluate whether models can select appropriate skills and tools, reason over intermediate observations and solve audio tasks through interaction.

\begin{figure*}[t]
\centering
\includegraphics[width=\textwidth]{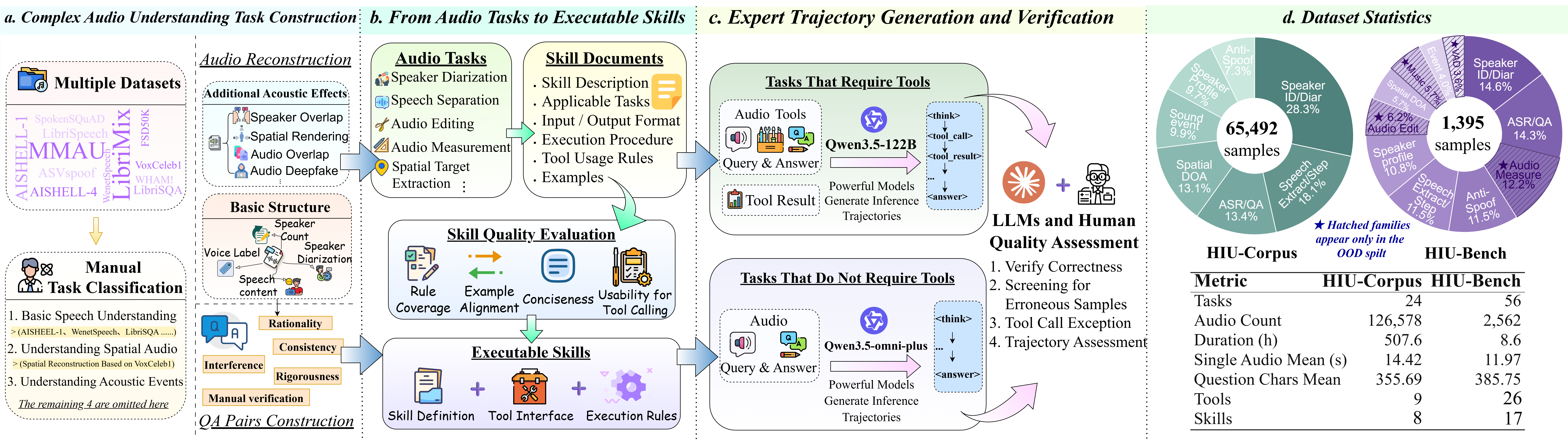}
\caption{Construction pipeline of HIU-Corpus and HIU-Bench, covering task collection, skill construction, trajectory generation and verification, and dataset statistics.}
\label{fig:dataset-construction}
\end{figure*}

\subsection{Audio and Voice Agents}

Recent studies have explored tool-based audio understanding from two directions. 
One line of work enhances audio reasoning through specialized tools, including model orchestration, tool routing, audio re-listening and evidence organization \cite{audiotoolagent,autagent,audiorouter,echo,audiomind,audiogenie,echoagent}. 
Although these methods improve audio reasoning, they treat capabilities in isolation rather than learning to dynamically coordinate acoustic skills based on intermediate observations. 
Another line of work develops general-purpose voice assistants that invoke external tools for digital tasks such as information retrieval, travel booking, and device control \cite{aura}, accompanied by benchmarks that evaluate tool selection, dependent API calls, multi-turn interaction and robustness under realistic speech conditions \cite{voiceagentbench,audio2tool,fullduplexbench}.
However, existing audio agents remain limited to isolated acoustic capabilities, while voice agents mainly treat speech as a command interface rather than audio evidence for reasoning.
In contrast, SpeechAgent-R learns multi-step coordination of acoustic skills and tools for complex audio problem solving, while HIU-Bench evaluates generalization to unseen task settings.

\section{Problem Formulation}
 
An audio agent should go beyond answering from fixed acoustic inputs: it should determine when additional skills are needed, retrieve relevant skills, invoke tools and use returned information to guide subsequent actions.
Given audio inputs $\mathcal{A}=\{A_1,A_2,\ldots,A_M\}$ and a user query $Q$, the agent has access to a user-defined skill library $\mathcal{S}=\{s_1,s_2,\ldots,s_N\}$. 
Each skill $s_i=(d_i,\mathcal{T}_i)$ consists of a natural-language description $d_i$ for skill selection and a set of associated tools $\mathcal{T}_i=\{t_{i1},t_{i2},\ldots,t_{iL_i}\}$. 
A tool processes input $x$ with parameters $\phi$ and returns observation $o=t(x,\phi)$, which can contain textual information or processed audio.
Rather than following a predefined workflow, the agent retrieves the corresponding skill document before invoking associated tools.
At step $k$, its context is defined as $h_k=(Q,\mathcal{A},\mathcal{D},a_{<k},o_{<k})$, where $\mathcal{D}$ denotes available skill information during interaction.
Based on $h_k$, the agent either retrieves the selected skill document, invokes a tool via action $a_k$, or terminates the interaction by generating response $y$. 
The resulting interaction forms a decision trajectory $\tau=\{(h_k,a_k,o_k)\}_{k=1}^{K}$. 
Since tool observations are produced by external tools, the agent learns to optimize its actions and final response as:
\begin{equation}
    (\hat{\tau},\hat{y})
    =
    \arg\max_{\tau,y}
    \left[
    \prod_{k=1}^{K}P_\theta(a_k\mid h_k)
    \right]
    P_\theta(y\mid h_{K+1}).
    \label{eq:problem_formulation}
\end{equation}
This formulation captures an adaptive \textit{Hear-Invoke-Understand} process, where agents select skills and tools based on intermediate evidence rather than fixed workflows.

\section{Method}


\subsection{HIU-Corpus Construction}
To provide diverse, high-quality trajectories for training generalizable audio agents, we follow three principles: 
tasks should i) require multi-step reasoning beyond direct audio understanding to capture meaningful decision processes; 
ii) combine multimodal understanding with external skills when needed and reason over their outputs rather than rely on tool execution alone; 
and iii) span diverse skills and workflows to encourage generalizable interaction beyond fixed workflows.
Guided by these principles, we construct HIU-Corpus through a three-stage pipeline (Fig.~\ref{fig:dataset-construction}).

\subsubsection{Complex Audio Understanding Task Construction}

Guided by these principles, we collect public datasets spanning speech recognition and audio QA \cite{wenetspeech,aishell-1,librisqa,mmau,spokensquad}, multi-speaker and mixed speech \cite{aishell-4,librimix}, speaker recognition and audio authenticity \cite{voxceleb1,asvspoof} and acoustic event understanding \cite{fsd50k}. 
We select 24 tasks that involve complex acoustic understanding, support reliable QA construction from source annotations, and cover scenarios where external skills may be required.
For each task, we manually design a template that maps source annotations to QA pairs and specifies the expected solution process, including whether external skills are needed. 
To increase acoustic complexity, we create additional acoustic scenes through audio mixing, temporal overlap and spatial rendering, while preserving task-relevant metadata such as speaker identities, timestamps and spatial positions.
We instantiate the templates using the original or constructed audio and use the original annotations as reference answers.
Finally, Claude-Opus-4.7 checks each QA pair against its template and annotations, with flagged samples manually corrected or removed.

\begin{figure*}[t]
\centering
\includegraphics[width=0.96\textwidth]{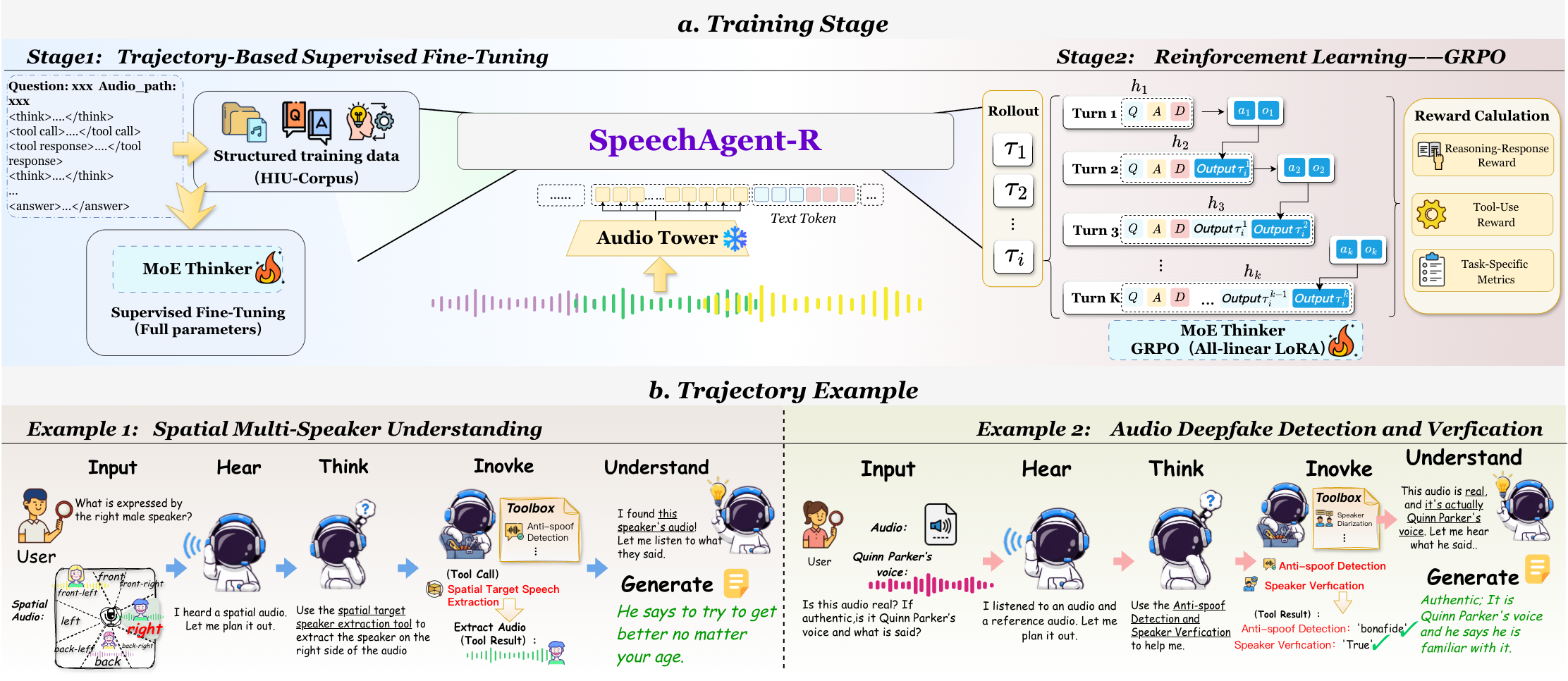}
\caption{Training and interaction workflow of SpeechAgent-R. 
(a) Two-stage training with trajectory-based SFT and multi-turn GRPO. 
(b) Example trajectories showing skill invocation and tool-output integration for two different tasks.}
\label{fig:training}
\end{figure*}

\subsubsection{From Audio Tasks to Executable Skills}
For tasks requiring external skills, related scenarios with similar problem-solving requirements are grouped into skills.
Each skill represents a task-oriented capability abstraction that organizes related tools, usage conditions and interaction patterns.
Within each skill, different scenarios may require different workflows, where one or more tools return text or processed audio for subsequent reasoning.
We therefore create a skill document for each skill, describing its tools, interfaces and calling rules, while providing guidance on when and how to select or combine tools across scenarios.
We then use Claude-Opus-4.7 to evaluate each document against real usage samples covering its tasks and major workflows, scoring interface accuracy, rule coverage, example alignment, conciseness and usability for tool calling.
Based on these evaluations, we manually refine the documents over multiple rounds.

\subsubsection{Trajectory Generation and Verification}

We construct direct-answer and skill-assisted trajectories following the interaction format used for agent execution.
For direct-answer tasks, Qwen3.5-omni-plus generates reasoning and responses from the audio, question and task prompt.
For skill-assisted tasks, we execute tools specified by the assigned skill, record their calls and outputs, and provide the QA instance, skill document and interactions to Qwen3.5-122B for trajectory generation.
Tool outputs may contain text or processed audio serve as observations for subsequent reasoning.
To prevent incorrect tool feedback from affecting SFT, we correct inaccurate observations using verified annotations while preserving the original tool-response format.
Claude-Opus-4.7 evaluates each trajectory for reasoning quality, decision quality and reasoning-answer consistency, with flagged samples manually corrected or removed.
Across skill and trajectory evaluations, scores range from 4.32 to 4.80 out of 5.
Within HIU-Corpus, 87.6\% of trajectories contain at least one tool call, and 10.7\% contain multiple calls.
Among tool-assisted trajectories, sequences average 3.24 turns (median 3, maximum 10), covering diverse tool-use patterns beyond direct audio answering.
HIU-Corpus is split into training and validation sets with a 95/5 ratio for training and model selection.

\subsection{SpeechAgent-R Training}
\textbf{Trajectory-based SFT } We initialize SpeechAgent-R from Qwen3-Omni-Thinking and perform SFT on verified interaction trajectories from HIU-Corpus to learn structured decision-making and tool-use behaviors. 
The sequence-modeling objective supervises only agent-generated tokens while masking user inputs and tool observations:
\begin{equation}
\mathcal{L}_{\mathrm{SFT}}
=
-\sum_{t=1}^{T}
m_t w_t
\log P_{\theta}(y_t\mid x,y_{<t})
\end{equation}
where $x$ denotes the input, $y=\{y_t\}_{t=1}^{T}$ the interaction sequence and $P_\theta$ the model probability. 
The mask $m_t$ excludes user and tool-observation tokens, while $w_t$ emphasizes agent actions such as skill selection and tool invocation.

\noindent\textbf{GRPO-based RL } We further optimize SpeechAgent-R with GRPO through multi-turn rollouts involving tool interaction. 
For each audio-query pair, the model samples a group of trajectories by selecting skills, invoking tools and reasoning over their outputs. 
We maximize the GRPO objective:
\begin{equation}
  \begin{gathered}
    \mathcal{J}_{\mathrm{GRPO}}(\theta)
    =\mathbb{E}\left[\frac{1}{G}\sum_{i=1}^{G}\frac{1}{T_i}\sum_{t=1}^{T_i}
    \left(\ell_{i,t}^{\mathrm{clip}}-\beta D_{\mathrm{KL}}^{i,t}\right)\right],\\
    \ell_{i,t}^{\mathrm{clip}}
    =\min\left(\rho_{i,t}\hat{A}_{i},\bar{\rho}_{i,t}\hat{A}_{i}\right).
  \end{gathered}
\end{equation}
where $G$ is the group size, $T_i$ is the number of model-generated tokens, $\rho_{i,t}$ is the current-to-old policy ratio, $\bar{\rho}_{i,t}$ is its $\epsilon$-clipped value, $\hat{A}_i$ is the group-relative advantage and $\beta$ is the KL weight.
We define the composite reward:
\begin{equation}
  R(\tau_i)=\lambda_fR_f(\tau_i)+\lambda_tR_t(\tau_i)+\lambda_aR_a(\tau_i).
\end{equation}
$R_f$ rewards valid reasoning-response structures and penalizes malformed or incomplete outputs. 
$R_t$ starts from 1 and penalizes missing or incorrect tools, redundant calls and execution errors. 
Both scores are clipped to $[0,1]$. 
$R_a$ uses normalized task-specific metrics, including EM for short answers, $1-\mathrm{CER}$ for ASR, BERTScore for open-ended responses, a combined transcription and turn-overlap score for diarization and path-set F1 for processed-audio outputs, with BLEU-4 and Token-F1 as fallback metrics.
For the final model, we set $(\lambda_f,\lambda_t,\lambda_a)=(0.05,0.25,0.70)$.
The composite rewards are then normalized within each group:
\begin{equation}
  \hat{A}_i=\frac{R(\tau_i)-\mu_R}{\sigma_R}
\end{equation}
where $\mu_R$ and $\sigma_R$ are the group reward mean and standard deviation. 
Because the reward is trajectory-level, $\hat{A}_i$ is shared across all model-generated tokens.
\section{HIU-Bench}

\begin{table*}[ht!]
\centering
\small
\setlength{\tabcolsep}{4pt}
\renewcommand{\arraystretch}{0.91}

\begin{tabular}{
    >{\raggedright\arraybackslash}
        m{\dimexpr 0.25\textwidth-2\tabcolsep\relax}
    *{10}{
        >{\centering\arraybackslash}
        m{\dimexpr 0.075\textwidth-2\tabcolsep\relax}
    }
}
\toprule
\multirow{2}{*}{\textbf{Model}}
& \multicolumn{4}{c}{\textbf{In-domain}}
& \multicolumn{4}{c}{\textbf{Out-of-domain}}
& \multicolumn{2}{c}{\textbf{Overall}} \\
\cmidrule(lr){2-5}
\cmidrule(lr){6-9}
\cmidrule(lr){10-11}
& Tool & Answer & Total & Time
& Tool & Answer & Total & Time
& Total & Time \\
\midrule

\multicolumn{11}{l}{
    \textit{(a) Direct setting}
} \\
\addlinespace[2pt]

Kimi-Audio
& -- & -- & 30.44 & 2.04
& -- & -- & 44.96 & 5.32
& 34.97 & 3.06 \\

Step-Audio-R1
& -- & -- & 37.99 & 11.16
& -- & -- & 24.86 & 11.39
& 33.90 & 11.23 \\

MiDashengLM
& -- & -- & 38.07 & 1.93
& -- & -- & 45.91 & 1.24
& 40.51 & 1.71 \\

Qwen2-Audio
& -- & -- & 37.78 & 1.07
& -- & -- & 39.65 & 0.73
& 38.36 & 0.96 \\

Qwen2.5-Omni
& -- & -- & 37.46 & 13.72
& -- & -- & 48.82 & 14.71
& 41.01 & 14.03 \\

Qwen3-Omni-Thinking
& -- & -- & 40.67 & 14.51
& -- & -- & 45.25 & 12.31
& 42.10 & 13.83 \\

Gemini-3.1-Pro-Preview
& -- & -- & 42.76 & 8.85
& -- & -- & 52.78 & 7.96
& 45.89 & 8.57 \\

Gemini-3-Flash-Preview
& -- & -- & 42.63 & 5.73
& -- & -- & 41.31 & 4.90
& 42.22 & 5.47 \\

\midrule

\multicolumn{11}{l}{
    \textit{(b) Agent harness setting}
} \\
\addlinespace[2pt]

Kimi-Audio
& 57.89 & 21.39 & 34.06 & 58.84
& 22.09 & 12.48 & 16.05 & 77.25
& 26.60 & 64.58 \\

Step-Audio-R1
& 79.97 & 49.26 & 58.90 & 25.32
& 32.78 & 15.86 & 22.14 & 24.82
& 46.24 & 25.16 \\

MiDashengLM
& 70.11 & 35.88 & 48.39 & 15.16
& 26.18 & 25.75 & 27.35 & 17.75
& 42.30 & 15.97 \\

Qwen2-Audio
& 71.25 & 44.66 & 53.17 & 1.50
& 31.95 & 32.31 & 33.10 & 0.79
& 47.18 & 1.28 \\

Qwen2.5-Omni
& 72.00 & 44.89 & 52.34 & 8.05
& 26.92 & 25.62 & 27.73 & 9.92
& 44.66 & 8.63 \\

Qwen3-Omni-Thinking
& 78.98 & 63.53 & 68.77 & 48.21
& 65.56 & 51.12 & 56.71 & 51.33
& 65.01 & 49.18 \\

Gemini-3.1-Pro-Preview
& 92.28 & 62.83 & 69.33 & 16.23
& 52.51 & 60.89 & 58.02 & 10.71
& 65.81 & 14.51 \\

Gemini-3-Flash-Preview
& 91.08 & 71.21 & 75.13 & 16.37
& 40.55 & 56.93 & 52.50 & 10.39
& 68.08 & 14.51 \\

\midrule

\rowcolor{gray!15}
SpeechAgent-SFT (ours)
& \textbf{97.94}
& \underline{70.74}
& \underline{78.92}
& 19.24
& \underline{78.51}
& \underline{60.56}
& \underline{66.98}
& 18.97
& \underline{75.20}
& 19.15 \\

\rowcolor{gray!15}
SpeechAgent-R (ours)
& \underline{96.74}
& \textbf{78.59}
& \textbf{84.17}
& 20.85
& \textbf{79.61}
& \textbf{65.80}
& \textbf{70.94}
& 25.45
& \textbf{80.05}
& 22.29 \\

\bottomrule
\end{tabular}

\caption{
Main results on HIU-Bench, reported separately on ID
and OOD splits.
Overall summarizes performance across both splits and all Time values are reported in seconds.
Boldface and underlining denote the best and
second-best results in each column, respectively.
Gray shading highlights our models.
}
\label{table-main}
\end{table*}












\begin{table*}[t]
\centering
\small
\setlength{\tabcolsep}{4pt}
\renewcommand{\arraystretch}{0.91}

\begin{tabular}{
    >{\raggedright\arraybackslash}
        p{\dimexpr 0.23\textwidth-2\tabcolsep\relax}
    *{7}{
        >{\centering\arraybackslash}
        p{\dimexpr 0.11\textwidth-2\tabcolsep\relax}
    }
}
\toprule
\multirow{2}{*}{\textbf{Setting}}
& \multicolumn{3}{c}{\textbf{In-domain (ID)}}
& \multicolumn{3}{c}{\textbf{Out-of-domain (OOD)}}
& \multirow{2}{*}{\textbf{Overall}} \\
\cmidrule(lr){2-4}
\cmidrule(lr){5-7}
& Tool & Answer & Total
& Tool & Answer & Total
& \\
\midrule

Base
& -- & -- & 40.67
& -- & -- & 45.25
& 42.10 \\

\textit{+ Agent harness}
& 78.98 & 63.53 & 68.77
& 65.56 & 51.12 & 56.71
& 65.01 \\

\midrule

\textit{+ SFT (answer-only)}
& 71.25 & 67.26 & 69.89
& 58.85 & 56.52 & 59.22
& 66.57 \\

\textit{+ SFT}
& \textbf{97.94} & \underline{70.74} & \underline{78.92}
& \underline{78.51} & \underline{60.56} & \underline{66.98}
& \underline{75.20} \\

\midrule

\rowcolor{gray!15}
\textit{+ SFT + RL (Ours)}
& \underline{96.74}
& \textbf{78.59}
& \textbf{84.17}
& \textbf{79.61}
& \textbf{65.80}
& \textbf{70.94}
& \textbf{80.05} \\

\bottomrule
\end{tabular}

\caption{Ablation of the agent framework and training strategies on
HIU-Bench. Best and second-best results are shown in bold and underlined,
respectively, gray denotes our final model.}
\label{tab:ablation_training}
\end{table*}

\subsection{Benchmark Construction and Statistics}

\begin{figure*}[]
\centering
\includegraphics[width=0.92\textwidth]{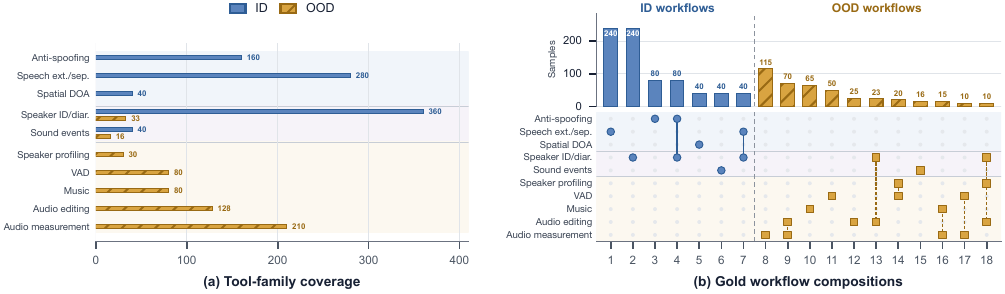}
\caption{Comparison of tool coverage and workflow compositions between ID and OOD splits in HIU-Bench. (a) Tool-family distribution. (b) Gold workflow compositions and their overlap across splits.}
\label{fig:id-ood}
\end{figure*}

HIU-Bench comprises an ID split that follows the same task, skill and tool distributions as HIU-Corpus while using independent audio instances, and an OOD split constructed through the same pipeline with new task scenarios and skill compositions.
It contains 1,395 samples across 56 tasks, including 960 samples from 24 ID tasks and 435 samples from 32 OOD tasks.
We check data independence between HIU-Corpus and HIU-Bench at the audio and task levels.
File-level matching and MD5 content verification find no binary-identical audio overlap in model inputs, while OOD tasks are disjoint from HIU-Corpus at the task level.
Speaker overlap in ID follows the splits of the source datasets, with no additional speaker leakage introduced during benchmark construction.
Furthermore, we compare tool coverage and workflow composition between ID and OOD.
Fig.~\ref{fig:id-ood}(a) shows different tool-family coverage, with OOD introducing additional tool categories, while Fig.~\ref{fig:id-ood}(b) shows 12 and 21 unique gold tool chains with no overlap.
These differences enable HIU-Bench to evaluate tool coordination across diverse task scenarios and workflow compositions.

\subsection{Evaluation Protocol}

We define $S_{\mathrm{total}}$ on HIU-Bench from three evaluation dimensions: format correctness ($S_f$), tool interaction ($S_t$) and answer quality ($S_a$), weighted by 0.05/0.25/0.70, with their effects analyzed in Section~\ref{experiment-ablation}.
For direct-answer evaluation, only $S_a$ is used. 
For agent evaluation, $S_f$ and $S_a$ follow the same criteria as $R_f$ and $R_a$, while $S_t$ is evaluated separately for ID and OOD tasks.
For ID tasks, $S_t$ follows the training-time scoring function, whereas OOD tasks adopt a workflow-aware rubric: expected tool usage receives full scores, valid alternatives partial scores, and invalid calls or failed executions limited scores.
Overall performance is computed as the sample-weighted average of ID and OOD scores.
The evaluation metrics follow the same task definitions as training rewards but are independently computed on held-out annotations, preventing direct reuse of training signals.

\section{Experiments}


\begin{table*}[t]
\centering
\small
\setlength{\tabcolsep}{4pt}
\renewcommand{\arraystretch}{0.91}

\begin{tabular}{
    >{\raggedright\arraybackslash}
        m{\dimexpr 0.34\textwidth-2\tabcolsep\relax}
    >{\raggedright\arraybackslash}
        m{\dimexpr 0.24\textwidth-2\tabcolsep\relax}
    *{3}{
        >{\centering\arraybackslash}
        m{\dimexpr 0.14\textwidth-2\tabcolsep\relax}
    }
}
\toprule
\textbf{Tool}
& \textbf{Metric}
& \textbf{Base}
& \textbf{SFT}
& \textbf{Ours} \\
\midrule

Speaker diarization
& $100 \cdot (1-\mathrm{DER})$
& 75.7 & 76.6 & 77.4 \\

Speaker verification
& Verified acc.
& 84.3 & 89.5 & 88.7 \\

Anti-spoof detection
& Label acc.
& 52.5 & 71.2 & 71.2 \\

Sound event detector
& Top-5 acc.
& 100.0 & 86.8 & 87.2 \\

Speech DOA estimator
& Acc.@$10^\circ$
& 100.0 & 100.0 & 100.0 \\

Target-speaker extraction
& SI-SDRi
& 61.9 & 62.2 & 62.2 \\

Spatial target-speech extraction
& SI-SDRi
& 26.9 & 32.0 & 41.6 \\

Speech separation
& SI-SDRi
& 74.5 & 77.4 & 75.9 \\

\midrule

\rowcolor{gray!15}
\textbf{Overall}
& {}
& \textbf{72.0}
& \textbf{74.5}
& \textbf{75.5} \\

\bottomrule
\end{tabular}

\caption{
Per-tool performance on the ID split of HIU-Bench.
Each tool is evaluated using the metric specified in the second column,
and Overall denotes the macro-average across all tools.
Higher values indicate better performance for all metrics.
}
\label{tab:tool_accuracy}
\end{table*}

\subsection{Experimental Setup}

SpeechAgent-R is initialized from Qwen3-Omni-Thinking and trained on 24 NVIDIA A100 GPUs using ms-swift in BF16. 
SFT fully fine-tunes the Thinker with a learning rate of $1\times10^{-6}$ and a global batch size of 32. 
For GRPO, we apply LoRA to all Thinker linear layers with rank 16 and alpha 32. 
The training uses a learning rate of $5\times10^{-7}$, a group size of 4, $\epsilon=0.2$, $\beta=0.04$, and reward weights $(0.05,0.25,0.70)$ for 150 steps.
The step-80 checkpoint is selected based on validation performance, and all models are evaluated under the same inference budget and agent environment.
We evaluate SpeechAgent-R on HIU-Bench against Kimi-Audio~\cite{kimi-audio}, Step-Audio-R1~\cite{step-audio-r1}, MiDashengLM~\cite{midashenglm}, Qwen2-Audio~\cite{qwen2-audio}, Qwen2.5-Omni~\cite{qwen3.5-omni}, Gemini-3.1-Pro-Preview and Gemini-3-Flash-Preview in direct and agent settings using the same scoring protocol.

\begin{figure*}[t]
    \centering
    \includegraphics[
        width=0.93\textwidth
    ]{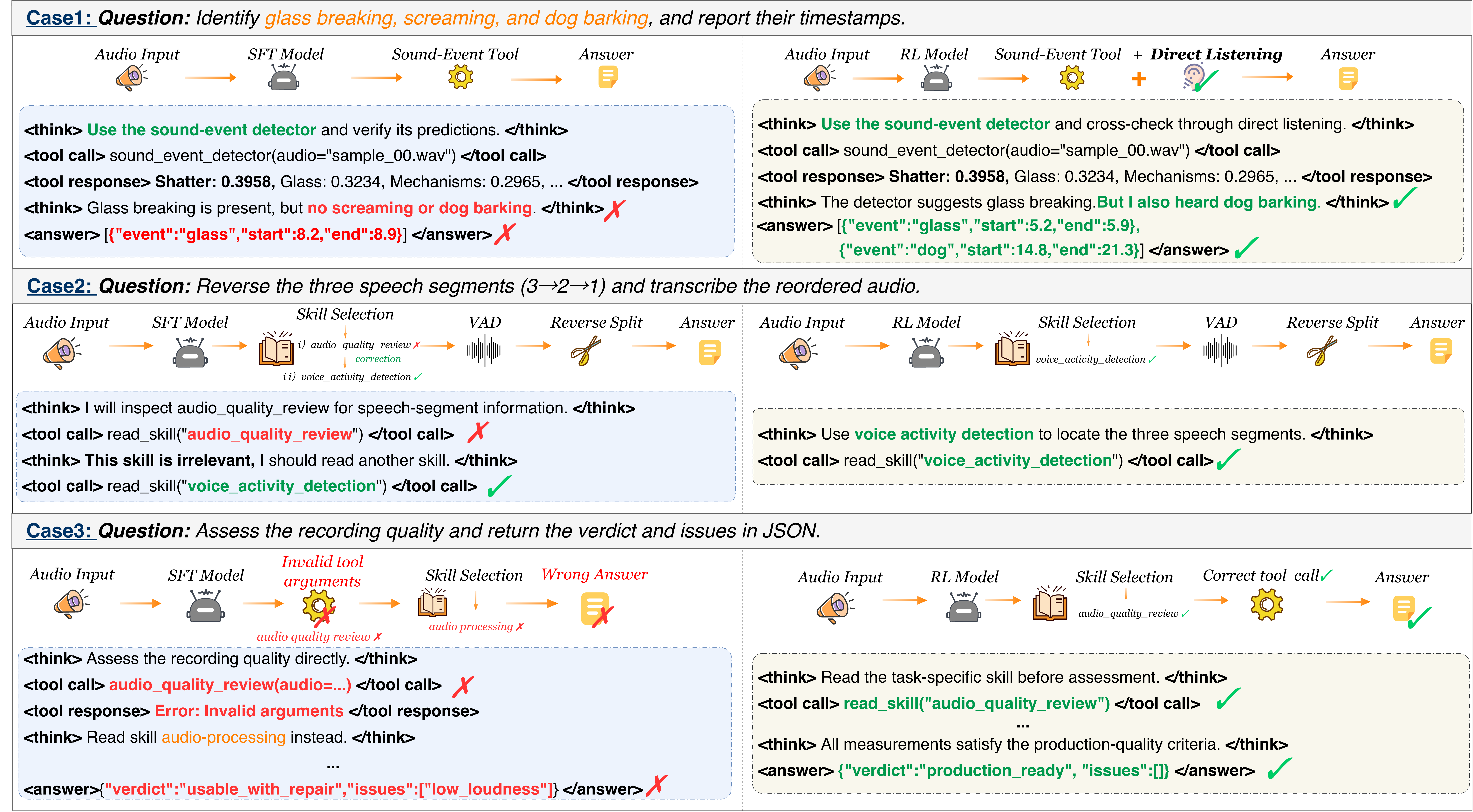}
    \caption{Representative cases comparing the trajectories generated by the SFT and RL models, illustrating differences in skill selection, tool usage and response generation.}
    \label{fig:case_study}
\end{figure*}

\subsection{Main Results on HIU-Bench}

Table~\ref{table-main} presents the results under two settings. 
Under the direct setting, all models achieve limited performance, with overall scores ranging from 33.90 to 45.89. 
These limitations are particularly evident in complex audio scenarios, such as spatial audio understanding and overlapping-audio analysis, where models must extract and integrate task-relevant acoustic evidence.
The shared agent harness improves most models, with Step-Audio-R1 rising from 33.90 to 46.24 and Gemini-3-Flash-Preview from 42.22 to 68.08.
However, the improvements vary across models: access to the same skills and tools does not lead to effective tool use by itself.
For example, Kimi-Audio achieves only 26.60 overall score, as unsuccessful tool interactions and format violations offset the benefits of tool access.
Moreover, some models also incur additional inference latency due to longer interaction trajectories. 
These results highlight the need for training audio agents that can effectively utilize skills and tools and generalize such capabilities to unseen scenarios.

After SFT on verified interaction trajectories from HIU-Corpus, SpeechAgent-SFT improves the overall score from 65.01 to 75.20 under the same agent harness, with gains on ID and OOD splits.
However, OOD analysis shows that SFT still faces tool-use challenges: 32.4\% of failures come from incorrect tool selection or poor tool-result use, and 18.2\% from ineffective interaction strategies.
With GRPO optimization, SpeechAgent-R achieves the highest overall score of 80.05, outperforming the base model with agent harness and SpeechAgent-SFT by 15.04 and 4.85 points.
OOD error analysis shows that GRPO improves agent behavior beyond supervised trajectories: compared with SFT, SpeechAgent-R reduces incorrect tool selection failures by 32.8\% and tool-result utilization failures by 16.3\%, enabling more effective skill use and response generation.

\subsection{Experiments Analysis}

\subsubsection{Ablation studies of different training strategies}
\label{experiment-ablation}

Table~\ref{tab:ablation_training} analyzes the effects of training strategies. 
Answer-only SFT, which supervises only final responses, achieves 66.57 overall, providing only a small improvement over the agent baseline, whereas trajectory-based SFT reaches 75.20 with gains on both ID and OOD tasks.
Building on SFT, GRPO further improves the score by 4.85 points to 80.05, showing that interaction trajectories establish agent behaviors and GRPO further refines skill selection and tool use. 
Reward weight ablations show that 0.05/0.25/0.70 weighting achieves the best overall score, while higher answer weights degrade tool-use performance.
Increasing the answer weight to 0.8 achieves 79.99 but causes more tool-use failures, highlighting the need to balance task performance and interaction reliability.

\subsubsection{Tool Interaction Analysis}

Table~\ref{tab:tool_accuracy} reports tool-level performance on the ID test set. 
Across different training strategies, tool execution performance remains comparable, with average scores of 72.0, 74.5, and 75.5 for Base, SFT, and SpeechAgent-R. 
These results suggest that the gains mainly come from improved agent-side skill and tool utilization rather than changes in tool capabilities.
To investigate the remaining performance bottleneck of SpeechAgent-R, we evaluate its upper bound under an oracle tool setting, where ground-truth tool outputs replace actual tool responses to eliminate tool execution errors. 
As shown in Table~\ref{tab:oracle_upperbound}, the overall score improves from 84.2 to 91.4, mainly driven by the increase in Answer score from 78.6 to 88.5, while Tool-call remains nearly unchanged. 
These results show that, after reliable skill selection and tool invocation are achieved, further improvements mainly depend on tool output quality and the model’s multimodal understanding and reasoning ability.

\subsubsection{Further Analysis}


\begin{table}[t]
\centering
\small
\setlength{\tabcolsep}{4pt}
\renewcommand{\arraystretch}{0.91}

\begin{tabular}{
    >{\raggedright\arraybackslash}
        m{\dimexpr 0.40\linewidth-2\tabcolsep\relax}
    *{3}{
        >{\centering\arraybackslash}
        m{\dimexpr 0.20\linewidth-2\tabcolsep\relax}
    }
}
\toprule
\textbf{Component}
& \textbf{Normal}
& \textbf{Oracle}
& $\Delta$ \\
\midrule

Format
& 99.5
& 99.3
& -0.2 \\

Tool-call
& 96.7
& 97.9
& +1.2 \\

Answer
& 78.6
& 88.5
& +9.9 \\

\midrule

\rowcolor{gray!15}
\textbf{Total}
& \textbf{84.2}
& \textbf{91.4}
& \textbf{+7.2} \\

\bottomrule
\end{tabular}

\caption{Oracle-tool analysis on the HIU-Bench ID split. $\Delta$ denotes Oracle minus Normal.}
\label{tab:oracle_upperbound}
\end{table}

Figure~\ref{fig:case_study} presents three representative cases showing how SpeechAgent-R improves tool interaction beyond SFT.
In Case 1, SpeechAgent-R improves tool-result utilization by integrating tool outputs with its own audio understanding, using multimodal evidence to refine predictions when tool responses are incomplete. 
In Case 2, SpeechAgent-R better understands task requirements and selects the appropriate skill, while SFT initially invokes an irrelevant skill and requires additional exploration. 
In Case 3, SpeechAgent-R further demonstrates skill-guided tool execution by understanding the relationship between skills and executable tools. 
It follows the skill requirements to invoke the necessary tools for the evaluation workflow, whereas SFT directly treats the skill name as an executable tool and produces an invalid tool call. 
Together, these cases show that SpeechAgent-R learns a reliable interaction process, covering skill selection, tool execution, and tool-result reasoning, which enables better generalization to complex audio tasks.
\section{Conclusion}

In this paper, we present SpeechAgent-R, a skill-calling multimodal agent that enables audio models to handle complex acoustic tasks through adaptive interaction. 
Through trajectory-based SFT on large-scale interaction data from HIU-Corpus and subsequent multi-turn RL, SpeechAgent-R learns to select skills, coordinate tools and use tool feedback for subsequent decisions beyond fixed-input audio understanding. 
We further introduce HIU-Bench to evaluate audio agents under both in-distribution and out-of-distribution settings.
\textit{SpeechAgent-R} achieves an overall score of 80.05 on HIU-Bench, improving over its base model under the same agent harness by 15.40 points and demonstrating improved skill and tool coordination across diverse task and workflow configurations.
One limitation is that, although HIU-Bench covers diverse task settings, extending it to more complex and realistic audio agent scenarios remains an important direction for future work. 
We will release the code, models, and datasets upon publication to support future research.

\bibliography{aaai2027}

\end{document}